\documentclass[aps,prl,preprint,groupedaddress]{revtex4}

\usepackage{epsfig}

\begin{document}

\title{Non-Destructive Identification of Cold and Extremely Localized Single Molecular Ions}

\author{M. Drewsen$^{\ast}$, A. Mortensen, R. Martinussen, P. Staanum, and J. L. S\o rensen}

\affiliation{Department of Physics and Astronomy, University of Aarhus \\
QUANTOP - Danish National Research Foundation Centre for Quantum Optics \\
Ny Munkegade, Bld. 520, DK-8000 Aarhus C, Denmark}

\date{\today}

\begin{abstract}

A simple and non-destructive method for identification of a single molecular ion sympathetically cooled by a single laser cooled atomic ion in a linear Paul trap is demonstrated. The technique is based on a precise determination of the molecular ion mass through 
a measurement of the eigenfrequency of a common motional mode of the two ions. The demonstrated mass resolution is sufficiently high that a particular molecular ion species can be distinguished from other equally charged atomic or molecular ions having the same total number of nucleons. 

\end{abstract}


\maketitle

In recent years, single molecule research has attracted massive attention. This is not at least due to its relevance for controlling processes in connection with nanotechnology and biophysics \cite{Gimzewski1999,Park2000,Kubatkin2003,Quake2000,Pascual2003,Weiss1999,Mehta1999,Gu2001,Moerner1999}. However, a single molecule represent as well an ultimate target for invetigations of many chemical physics processes. A single molecular ion sympathetically cooled \cite{Larson1986} through the Coulomb interaction with a single laser cooled atomic ion in a linear Paul trap \cite{Prestage1989} represents a translationally cold ($T<$10 mK) and spatially extremely well-localized target(occupied volume: $\sim$ 10 $\mu$m$^{3}$) for such single molecule investigations. In contrast to the situation in many recent investigations of neutral single molecules \cite{Gimzewski1999,Park2000,Kubatkin2003,Quake2000,Pascual2003,Weiss1999,Mehta1999,Gu2001,Moerner1999} where the molecule is either adsorbed on surfaces \cite{Gimzewski1999,Park2000,Kubatkin2003,Pascual2003}, contained in a fluid \cite{Weiss1999,Mehta1999,Gu2001} or embedded in a host material \cite{Moerner1999}, a single trapped molecular ion can essentially be free from perturbations from its environment. This makes it possible to perform, e.g., well-controlled reaction experiments \cite{Moelhave2000,Drewsen2002} and studies of quantum coherence in photo-ionization and -dissociation processes \cite{Shapiro2000,Brixner2001}. The long perturbation free trapping times (typically of the order of minutes) make it furthermore realistic to consider laser cooling schemes for the internal degrees of freedom of the molecules\cite{Vogelius2002}, a first step towards state selective single molecule studies. A key issue in all such experiments is to identify the trapped molecular ion species.      

In this Letter, we demonstrate a simple technique to uniquely identify a single trapped molecular ion which relies on a non-destructive measurement of its mass through resonant excitation of an oscillation mode of the two-ion system. In the past, the simple relation between masses and oscillation frequencies has been exploited in a large variety of measurements, including the most precise mass measurement of molecular ions today \cite{DiFilippo1994,mit1999,Rainville2004}. In contrast to these precision experiments, a strong Coulomb coupling between the molecular ion of interest and an additional atomic ion is essential in our identification scheme. For the harmonic potential of our trap, one can readily show that for two equally charged ions close to their equilibrium positions, the two eigenfrequencies for their motion along the axis of alignment are given by \cite{Kielpinski2000}

\begin{equation}
\label{modefreqs} 
\nu_{+/-}^2=\Bigl[ \left( 1+\mu \right)\pm\sqrt{1-\mu+\mu^2} \Bigr] \nu_{1}^2, 
\end{equation}
with  $\mu= M_{1}/ M_{2}$, where $M_{1}$ and $M_{2}$ are the masses of the two ions, and $\nu_{1}$ is the oscillation frequency of a single ion with mass $M_{1}$.  In the following, the corresponding motional modes will  be referred to as the center-of-mass mode ($\nu_{-}$) and the breathing mode ($\nu_{+}$). By determining  $\nu_{1}$ for an ion with a known mass $M_{1}$, the mass $M_{2}$ of the other ion can be deduced from a measurement of one of the frequencies $\nu_{+}$ and $\nu_{-}$. 
For two identical ions with mass $M_{1}$, the center-of-mass mode frequency is equal to $\nu_{1}$. Since two equally charged ions have identical equilibrium positions independent of their masses, it is hence possible to measure $\nu_{1}$ and $\nu_{-}$ with ions located at the same positions in the trap, which minimizes systematic errors arising from slight anharmonicities in the trap potential. 

The essential parts of our experiment are sketched in Fig. 1. The Ca$^+$ ions used in the experiments to sympathetically cool the molecular ions were produced by an isotope selective photoionization process \cite{Kjaergaard2000,Mortensen2004}, while the molecular ions used for illustrating the technique were produced in reactions between the Ca$^+$ ions and a thermal gas of O$_{2}$ \cite{Drewsen2002}. In the experiments, we apply laser cooling to at least one $^{40}$Ca$^{+}$ ion, such that the ions are directly or sympathetically cooled down to temperatures of a few mK. At such low temperatures, the ions are localized to within a small fraction of their equilibrium distance \cite{Rohde2001,Drewsen2002}. The ions are forced to line up along the trap axis by making the radial trap frequencies higher than the axial one. The positions of the $^{40}$Ca$^{+}$ ions are directly be observed by a CCD-camera through detection of fluorescence light emitted during the laser cooling process. An image of two $^{40}$Ca$^{+}$ ions is shown in Fig. 2(a). The two-ion eigenfrequencies are measured by applying a harmonically oscillating perturbing force along the alignment axis. When the drive frequency of the force is near one of the eigenfrequencies, the amplitude of the ions's motion along this axis is enhanced, which for exposure times longer than the oscillation period leads to a spatial smearing of the fluorescence signal in the CCD-images as shown in Fig. 2(b). In the experiments we have applied a perturbing force in two different ways. The simplest method consists of applying an oscillating electric field along the trap axis by adding an oscillating voltage to two diagonally positioned end electrodes as shown in Fig. 1. The other method is based on modulating the radiation pressure force on the laser cooled ion(s) by a modulation of the laser intensity of one of the beams propagating along the trap axis (see Fig. 1). 

The images presented in Fig. 2 show snapshots from a series of measurements where $\nu_{1}$ was measured for two $^{40}$Ca$^{+}$ ions and $\nu_{-}$ for one $^{40}$Ca$^{+}$ ion and one unknown ion, which presumably is a $^{40}$Ca$^{16}$O$^{+}$ ion formed in a reaction between one of the ions shown in Fig. 2(a) and an O$_{2}$ molecule. Figs. 2(a) and 2(c) show the fluorescence from the laser cooled $^{40}$Ca$^{+}$ ions before any perturbing force was applied. Since the single $^{40}$Ca$^{+}$ ion in Fig. 2(c) is located exactly at the same position as one of the ions in Fig. 2(a), we can be confident that there is only one non-visible singly charged ion present. When tuning the frequency of a perturbing electric field to maximize the amplitude of the ions's forced motion for the two cases presented in Fig. 2, we found the resonance frequencies $\nu_{1}=98.7(1)$ kHz and $\nu_{-}=89.4(1)$ kHz, respectively. The mass $M_{1}$ of the $^{40}$Ca$^{+}$ ion is known to be 39.97 a.m.u., and by combining this with the found frequencies via Eq. \ref{modefreqs} one gets $M_{2}=56.1(4)$ a.m.u. Although not impressive, this precision is already sufficient to conclude that the non-visible molecular ion is indeed a $^{40}$Ca$^{16}$O$^{+}$ ion and not a CaO$^+$ ion with another isotopic composition.

To discriminate between equally charged atomic or molecular ions with the same total number of nucleons, a relative mass resolution of $\sim 10^{-4}$ is typically needed. This level of accuracy can not be reached using the simple technique discussed above, since the measured resonance frequency is subject to a systematic shift of up to $\sim$1\%, depending on the damping effect of the  laser cooling force and the trap frequency. However, by detecting only the in-phase component of the motion of the ions with respect to the periodic perturbing force, a dispersive shaped signature of the trap resonance is found. The zero crossing of this curve yields the trap resonance frequency free of lowest order damping dependent shifts. In this manner a relative mass resolution better than $\sim 10^{-4}$ is indeed feasible\cite{Soerensen2004}. To demonstrate such a mass resolution, we have chosen to work with ions of different isotopes of calcium as test masses instead of molecular ions, since in our setup it is relatively easy to create ions of the various isotopes either by isotope selective photoionization\cite{Kjaergaard2000,Mortensen2004} or charge exchange processes\cite{Mortensen2004, Drewsen2003}. In Fig. 3, the position resolved fluorescence of two $^{40}$Ca$^{+}$ ions (Fig. 3(a)) and one $^{40}$Ca$^{+}$ ion and one $^{42}$Ca$^{+}$ ion (Fig. 3(b)) is shown as a function of the intensity modulation frequency of a single cooling laser beam. For these data the laser intensity was modulated with the laser field only present for 1.3 $\mu$s in each period, and with the CCD-camera gated to only record light emitted within a time span of 1.1 $\mu$s during the presence of the modulated light. In both the cases presented in Fig. 3, the phase of the ions's motion changes rapidly around a specific modulation frequency. With $M_{1}$ being the mass of the $^{40}$Ca$^{+}$ ion, according to the discusssion above these frequencies equal the resonance frequencies $\nu_{1}$ and $\nu_{-}$. The resonance frequencies are derived from fits to data series like those shown in Fig. 3, where the spatial fluorescence intensity distribution of one ion is modelled by the
function $I(z,\nu_{drive})=I_0
\exp\left\lbrace-\frac{[z-z_0(\nu_{drive})]^2}{\Gamma_z(\nu_{drive})
^2}\right\rbrace+I_B$, where $\nu_{drive}$ is the drive frequency, $I_0$ is the ion signal, $I_B$ is the background level,
$z_0(\nu_{drive})$ is the mean in-phase position of the ion and $\Gamma_z
(\nu_{drive})=\frac{\alpha_\nu}{(\nu_{drive}-\nu_-)^2+\gamma_\nu^2}+\Gamma_{res}$ is the spatial width of the fluorescence signal with $\Gamma_{res}$ being the spatial resolution of the imaging
system, $\gamma_\nu$ being the width of the resonance and $\alpha_\nu$ being a constant related to the amplitude of the forced motion. Finally, $z_0 (\nu_{drive})=
\frac{\alpha_z(\nu_{drive}-\nu_-)}{(\nu_{drive}-\nu_-)^2+\gamma_\nu^2}+z_{off}$,
where $z_{off}$ is the ion position in the absence of modulation
forces and $\alpha_z$ is a constant related to the amplitude of the forced
motion. In Fig. 4, the results of such fits are presented for three combinations of calcium isotopes. From a statistical analysis of these data, we find $\mu_{40/42, meas}=0.9526(3)$ and $\mu_{40/44, meas}=0.9095(3)$, respectively. Using the best known values for the atomic masses of the calcium isotopes available \cite{Audi1993} and subtracting the mass of one electron, the corresponding expected values are $\mu_{40/42, table}=0.952441$ and $\mu_{40/44, table}=0.909174$, respectively. We see that our relative mass measurements coincide with the expected ones at a level of $1.4\times 10^{-4}$ and $3.4\times 10^{-4}$, respectively. These deviations reflect the size of the systematic errors in the experiments due to the asymmetry, which arises when applying laser cooling to only one ion, as well as the timing resolution of the camera system \cite{Soerensen2004}. The accuracy in these measurements is already sufficient to discriminate between some singly charged atoms or molecules with an identical number of nucleons, such as $^{13}$C$_2$H$_2^+$ and $^{14}$N$_2^+$ which have a relative mass difference of $5.8\times10^{-4}$ \cite{Rainville2004}. Without significant changes in our setup, we should in the future be able to reach the level of $10^{-5}$ or lower \cite{Soerensen2004}. This level of accuracy can of course not compete with the precision of the best dedicated mass measurements scheme \cite{Rainville2004}, but it will certainly be sufficient for most chemical physics experiments.

Since the presented identification technique does not rely on spectroscopic sideband resolution, it can be applicable in a large variety of experiments. The range of the molecular mass-to-charge ratio, where the technique can be used, spans at least the range from 2 to $\sim 500$, depending on the choice of the laser cooled ion \cite{Ions}. As already indicated by the results in Fig. 2, the identification technique is an excellent tool in single ion reaction experiments, and its non-destructive nature as well as the short measurement time needed ($\sim$ 1 min) makes it particularly useful when consecutive reactions \cite{Drewsen2002} or other successive manipulations are pursued. Since processes leading to a change in the mass or charge of the target molecule can be detected with essentially 100 $\%$ efficiency, the technique is very attractive for, e.g., single molecule photo-ionization or -dissociation studies. This includes coherent-control experiments \cite{Shapiro2000,Brixner2001} with the molecular ion prepared in a specific internal state, e.g., by cooling techniques \cite{Vogelius2002}.  
It should be noted that, in contrast to the electrical perturbation scheme\cite{Naegerl1998}, the laser force modulating technique allows easy excitation of both the center-of-mass mode and the breathing mode, such that both the frequencies $\nu_{-}$ and $\nu_{+}$ can be determined. Knowing these two frequencies, no reference measurement of $\nu_{1}$ is needed in order to determine $M_2$. 

In conclusion, we have presented a simple single molecule identification technique which allows a non-destructive discrimination between equally charged atomic and molecular ions with an identical number of nucleons. Through the sympathetical cooling procedure a localized and cold single molecule target is available for further studies in an environment with very weak perturbations.
\begin{acknowledgments}
This work was supported by the Danish National Research Foundation.
\end{acknowledgments}

\bibliography{singlebib}

\begin{thebibliography}{28}
\bibitem[{Ion()}]{Ions}
\bibinfo{note}{This range is valid when more laser coolable atomic ions are
  available (e.g.,$^{9}Be^+$,$^{40}Ca^+$ and $^{138}Ba^+$). For a specific
  choice of coolant ion the range is smaller.}
\end{thebibliography}

\bibliographystyle{apsrev}
\begin{figure}[p]
  \centering
\includegraphics[width=\textwidth]{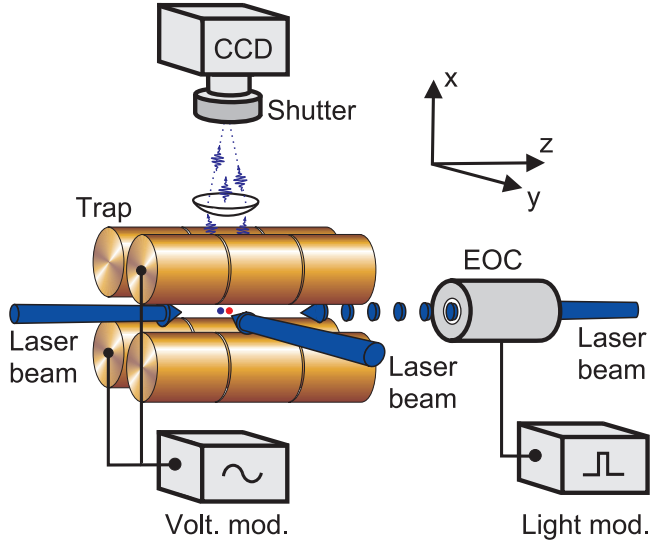}\\\vspace{-6mm}
  \caption{(Color online) Sketch of the experimental setup. The applied linear Paul trap (Trap) gives rise to an effective three-dimensional harmonic potential for the ions (see Ref.\cite{Drewsen2003}). By making the oscillation frequency along the trap axis ($z$-axis) lower than the trap frequencies in the $x-y$ plane, two cold ions will line up as indicated by the two dots in the figure. Cooling of the $^{40}$Ca$^+$ ion (left (blue) dot) is provided by three laser beams as shown, while the molecular ion (right (red) dot) will be cooled sympathetically. The forced motion of the ions is induced either by applying a sinusoidally varying voltage to two end-electrodes of the trap or by modulating the radiation pressure force from one of the laser beams propagating along the trap axis by chopping its intensity using an electrooptical chopper (EOC). The fluorescence light from the $^{40}$Ca$^+$ ion is imaged onto a CCD-chip through a fast electronic shutter, which can be gated synchroneously with the phase of the applied modulated force.}
\label{Fig1}
\end{figure}

\begin{figure}[p]
\centering
\includegraphics[width=\textwidth]{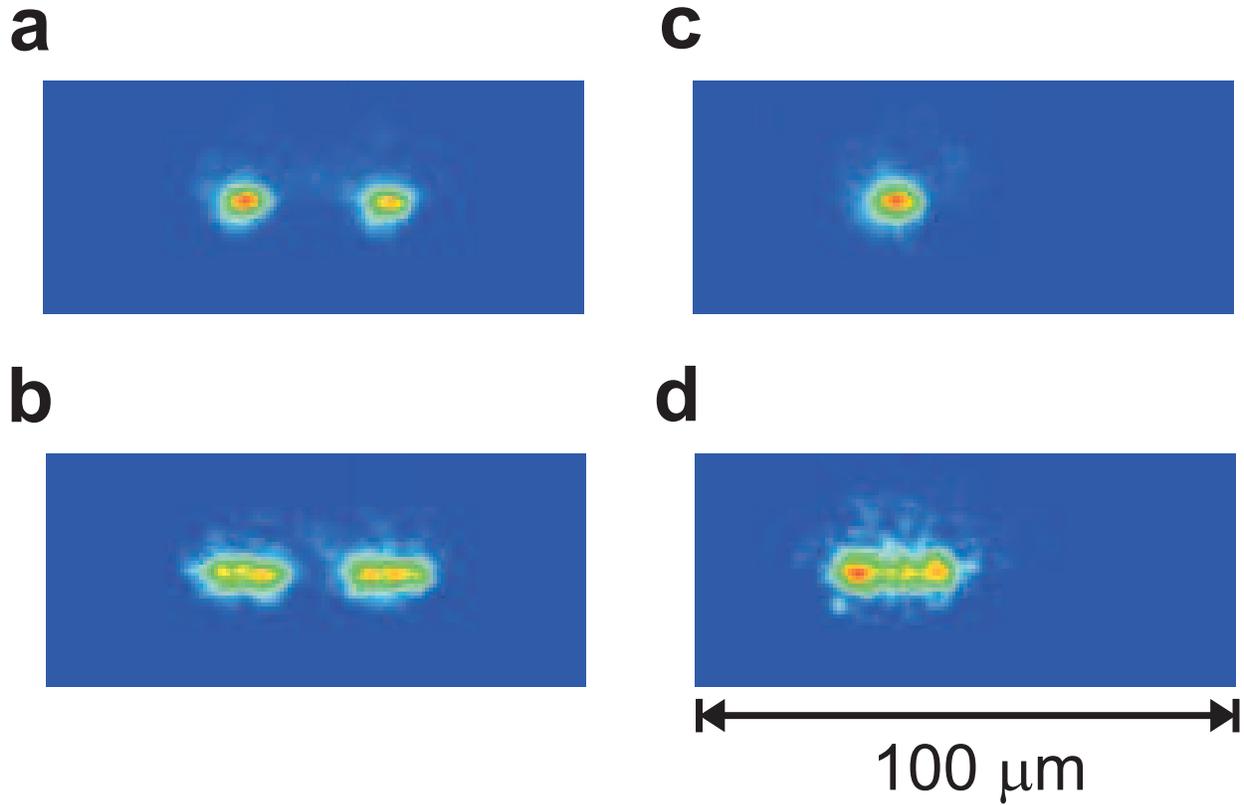}\\\vspace{-6mm}
  \caption{(Color online) Images of fluorescence from $^{40}$Ca$^+$ ions. (a) Two $^{40}$Ca$^+$ ions at thermal equilibrium in the trap. (b) The same two $^{40}$Ca$^+$ ions as in (a), but with a modulated voltage applied to the trap electrodes at a frequency near the resonance frequency $\nu_{1}$=98.7 kHz. (c) The same situation as in (a), but after one of the $^{40}$Ca$^+$ ions had reacted with an O$_2$ molecule and presumably formed a non-fluorescing $^{40}$Ca$^{16}$O$^{+}$ ion. (d) Image of the $^{40}$Ca$^+$ ion in (c), but when a modulated force at a frequency close to the resonance frequency $\nu_{-}$=89.4 kHz is present. In all the experiments the radial trap frequencies were 380 kHz and the exposure time of the CCD-chip was 100 ms.
}
\label{Fig2}
\end{figure}

\begin{figure}[p]
  \centering
\includegraphics[width=\textwidth]{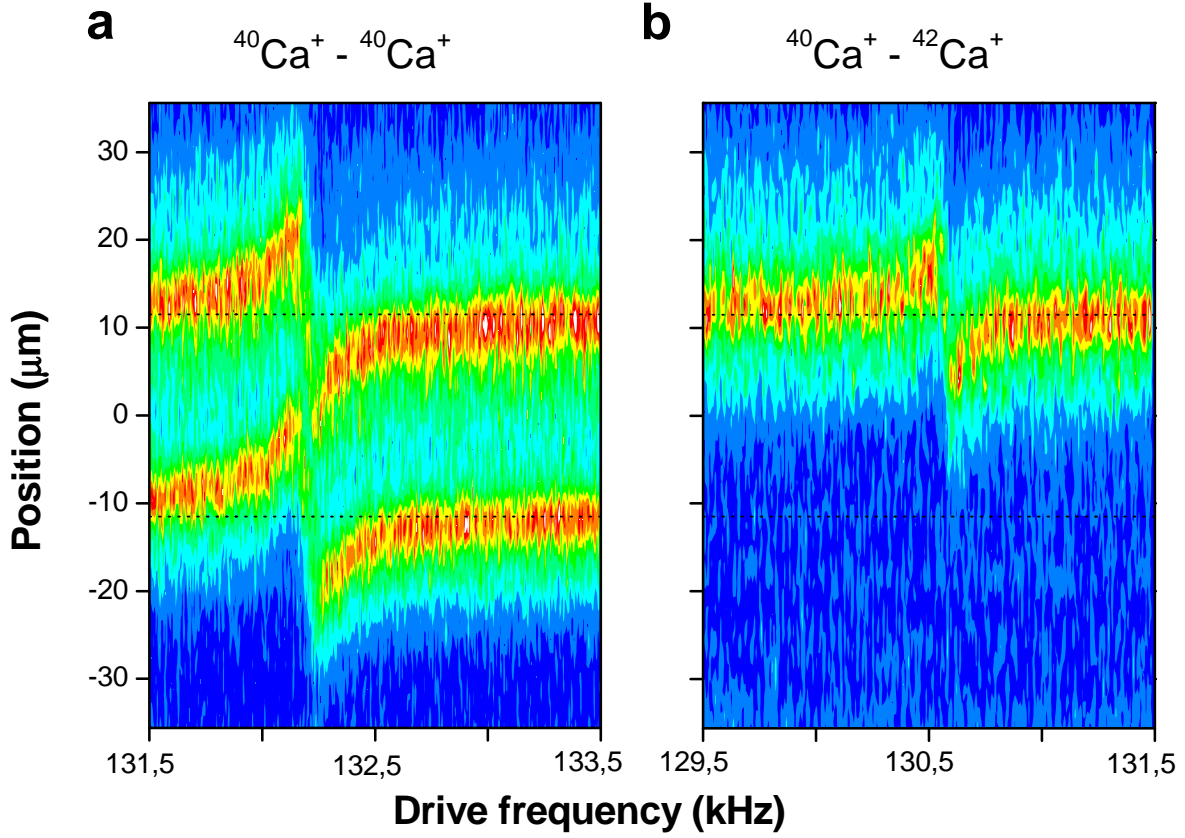}\\\vspace{-6mm}
  \caption{(Color online) The position resolved fluorescence along the trap axis as a function of the drive frequency of the intensity modulated laser beam. (a) Two $^{40}$Ca$^{+}$ ions. (b) One $^{40}$Ca$^{+}$ ion and one $^{42}$Ca$^{+}$ ion, where only the $^{40}$Ca$^{+}$ ion is fluorescing. Each false-colored contour plot is composed of axial projections of the fluorescence intensities in gated images recorded during the frequency scans (Scan rates: (a) 2 Hz/image, (b) 8 Hz/image). Red (blue) indicates high (low) fluorescence level. The dashed lines in the figures indicate the equilibrium positions of the ions when no modulated force is present. The radial trap frequencies were 400 kHz in all experiments.
}
\label{Fig3}
\end{figure}

\begin{figure}[p]
  \centering
\includegraphics[width=\textwidth]{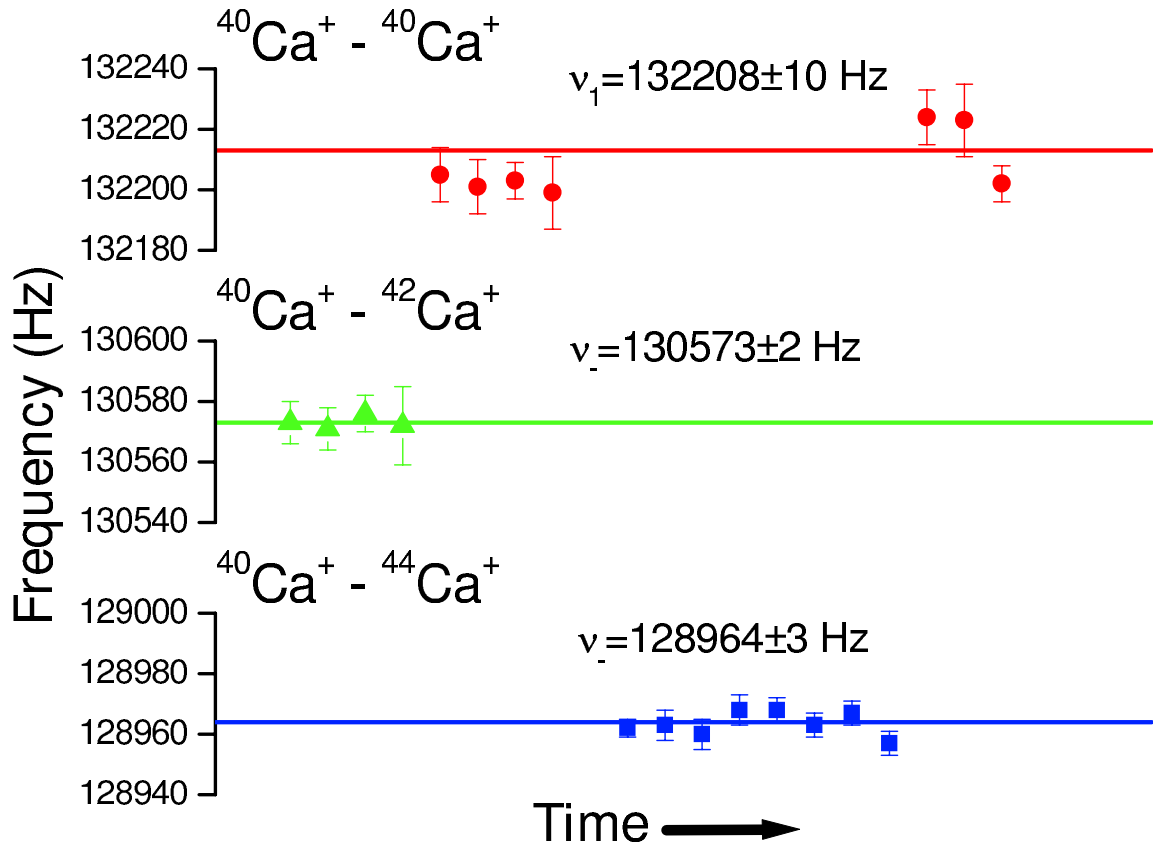}\\\vspace{-6mm}
  \caption{(Color online) Resonance frequencies obtained for three combinations of calcium isotopes. Each data point represent the resonance frequency found from fits to data series as those shown in Fig. 3. The stated resonance frequencies present the statistical average values of the data points. The horizontal position of the data points reflects the time order of the experiments.
}
\label{Fig4}
\end{figure}

\end{document}